# MATAI: A Generalist Machine Learning Framework for Property Prediction and Inverse Design of Advanced Alloys

Yanchen Deng[1], Chendong Zhao[2], Yixuan Li[1], Bijun Tang[2✉], Xinrun Wang[3✉], Zhonghan Zhang[2], Yuhao Lu[1], Penghui Yang[1], Jianguo Huang[1], Yushan Xiao[2], Cuntai Guan[1✉], Zheng Liu[2,4,5✉], Bo An[1✉]

[1]College of Computing and Data Science, Nanyang Technological University, Singapore

[2]School of Materials Science and Engineering, Nanyang Technological University, Singapore

[3]School of Computing and Information Systems, Singapore Management University, Singapore

[4]CINTRA CNRS/NTU/THALES, UMI 3288, Research Techno Plaza, 50 Nanyang Drive, Border X Block, Level 6, Singapore 637553, Singapore

[5]Institute for Functional Intelligent Materials, National University of Singapore, Singapore, Singapore

[6]These authors contributed equally: Yanchen Deng, Chendong Zhao, Yixuan Li

✉E-mail: bjtang@ntu.edu.sg; xrwang@smu.edu.sg; ctguan@ntu.edu.sg; z.liu@ntu.edu.sg; boan@ntu.edu.sg

The discovery of advanced metallic alloys is hindered by vast composition spaces, competing property objectives, and real-world constraints on manufacturability. Here we introduce MATAI, a generalist machine learning framework for property prediction and inverse design of as-cast alloys. MATAI integrates a curated alloy database, deep neural network-based property predictors, a constraint-aware optimization engine, and an iterative AI–experiment feedback loop. The framework estimates key mechanical properties—including density, yield strength, ultimate tensile strength, and elongation—directly from composition, using multi-task learning and physics-informed inductive biases. Alloy design is framed as a constrained optimization problem and solved using a bi-level approach that combines local search with symbolic constraint programming. We demonstrate MATAI's capabilities on the Ti–based alloy system, a canonical class of lightweight structural materials, where it rapidly identifies candidates that simultaneously achieve lower density (<4.45 g/cm$^3$), higher strength (>1000 MPa) and appreciable ductility (>5%) through only seven iterations. Experimental validation confirms that MATAI-designed alloys outperform commercial references such as TC4, highlighting the framework's potential to accelerate the discovery of lightweight, high-performance materials under real-world design constraints.

## 1. Introduction

Alloys are the backbone of modern materials engineering, offering a broad range of tunable properties including mechanical strength, fatigue and wear resistance, thermal and corrosion stability, saturation magnetization, and electrical conductivity, which enable their use in an exceptionally wide array of applications[1,2]. From structural components in aerospace and automotive systems to functional materials in precision electronics and energy infrastructure for sustainable energy supply, alloys are indispensable in advancing technological progress across industries[3,4]. This versatility arises from their compositional flexibility and the complex interactions among constituent elements, which allow fine-tuning of material properties to meet targeted performance requirements[5,6]. Despite these advantages, alloy discovery remains a formidable challenge. Even when considering 20 commonly used metallic elements, the number of possible equiatomic five-component combinations exceeds 15,000, and this number grows exponentially with the inclusion of variable stoichiometries, processing conditions, and target properties[5,7,8]. The resulting design space is vast, highly non-linear, and largely underexplored. Consequently, alloy development continues to rely heavily on expert heuristics and trial-and-error experimentation, often requiring years of iterative testing to yield viable candidates, rendering the process both time-consuming and prohibitively expensive[9].

Machine learning (ML) has recently emerged as a promising data-driven approach for accelerating materials discovery, including alloy design. By learning complex, nonlinear relationships between composition and properties from historical data, ML has enabled rapid surrogate modelling, property prediction, and even inverse design, reducing the need for costly simulations and extensive physical screening[10]. shown particular promise in high-entropy alloy (HEA) research, aiding the identification of candidate systems for both mechanical and functional applications[11,12]. However, existing ML approaches remain constrained in their scope and generalizability. Most models are trained on narrow, system-specific datasets and are typically tailored to specific target properties, such as phase classification[13], hardness[14], or adsorption energy[15], limiting their transferability to new systems without significant retraining. Additionally, current approaches usually treat the learned model as a high-throughput screening tool, either relying on expert-proposed candidates or performing exhaustive searches within a limited compositional subspaces[11,16]. While some studies[12,17] incorporate optimization strategies such as genetic algorithm (GA)[18] or Markov chain Monte Carlo (MCMC) sampling[19] to explore broader design spaces, they often overlook real-world hard constraints, such as upper limits on the number of active elements or maximum allowable concentrations of minor elements, which are essential for practical manufacturability. Critically, most methods[11,12,17] optimize only a single target property, despite alloy design inherently involving trade-offs between multiple, often conflicting, objectives.

In light of this, we introduce MATAI, a generalist ML framework for alloy property prediction and inverse design. Unlike task-specific models, MATAI integrates domain knowledge from diverse alloy systems and supports multi-objective, constraint-aware optimization across broad compositional spaces. The framework consists of four core components: 1) a holistic alloy database containing over 10,000 experimentally verified compositions, aggregated from open databases, literature, and in-house experiments; 2) foundational property predictors capable of estimating multiple alloy properties such as density, yield strength (YS), ultimate tensile strength (UTS), and elongation directly from alloy compositions; 3) a generalist alloy designer that performs constrained optimization over multiple objectives, enabling the discovery of promising alloy candidates without exhaustive searches; and 4) an iterative AI-experiment feedback loop that continuously refines the model through experimental validation of AI-generated candidates.

To demonstrate the effectiveness and robustness of MATAI, we apply the framework to the titanium (Ti)-based alloys, a canonical aerospace alloy system valued for its low density with high strength. Using MATAI, we identified novel compositions that achieve high strength (>1000 MPa) and moderate elongation (>5%) while retaining a low density (<4.45 g/cm$^3$), surpassing commercial benchmarks. These results highlight MATAI's potential to accelerate the discovery of high-performance alloys under realistic design constraints through intelligent, data-driven exploration.

## 2. Results

### 2.1 Overview of the MATAI Framework

MATAI is a modular ML framework designed to accelerate general-purpose alloy discovery and optimization through a closed-loop neuro-symbolic architecture[20]. It integrates four key components: a holistic alloy database, a set of foundational property predictors, a constraint-aware generalist designer, and an iterative AI-experiment feedback loop (Figure 1). Together, these components form a unified pipeline for data-driven inverse design under realistic material constraints. The framework begins with the construction of a consolidated alloy database (Figure 1a) that compiles data from public databases, literature, and in-house experiments into a standardized format. This database provides the training foundation for MATAI's deep neural network-based foundational predictors, which estimate key mechanical properties such as density, YS, UTS, and elongation directly from composition.

These predictors (Figure 1b) serve as high-fidelity surrogates for property evaluation, enabling high-throughput property estimation across large compositional spaces. Guided by these models, the generalist alloy designer (Figure 1c) formulates the design task as a discrete constrained optimization problem (COP)[21,22], incorporating both hard physical rules and soft user-defined preferences. The designer solves this problem using a bi-level optimization strategy (Figure 1d) that combines stochastic local search[23,24] with symbolic constraint solving[25,26].

Top-ranked alloy candidates generated by the designer are synthesized and experimentally evaluated. Results, both successful and failed, are fed back into the dataset via an iterative AI–experiment feedback loop (Figure 1e), allowing the predictors to be retrained and the designer to explore increasingly refined solution spaces over successive iterations. This iterative learning process improves both prediction accuracy and design quality over time. Although MATAI is broadly applicable across alloy systems, we demonstrate its effectiveness in this study through a focused application: the design of Ti-based lightweight alloys that simultaneously deliver high specific strength at lower density while preserving appreciable ductility.

## 2.2. Construction of Holistic Alloy Database

To support generalizable ML models for alloy property prediction and inverse design, we constructed a comprehensive and high-quality dataset by integrating data from six distinct sources: four publicly available databases (Multi-Principal Element Alloys (MPEA)[27], MakeItFrom, Mg-based[28,29], and Al-based[30]), a manually curated literature dataset verified by domain experts, and in-house experimental results from MATAI-designed alloys. Each source was selected to maximize diversity, reliability, and representativeness across alloy families and property target. This unified database provides the training foundation for MATAI's foundational predictors and generalist designer.

The dataset spans a wide range of alloy families and includes both common and niche alloying elements. Figure 2a presents the elemental distribution across the periodic table, with shading indicating the frequency of appearance in the dataset. High-frequency elements such as Fe, Al, Ni, and Mn reflect their widespread use in structural alloys, while lighter representation of others (e.g., rare earths) supports compositional diversity beyond conventional systems. To assess coverage in the high-dimensional compositional space, we apply t-SNE embedding to the entire dataset (Figure 2b). The resulting 2D projection reveals that data from sources such as MPEA, MakeItFrom, Al-, Mg-based alloys, literature, and our own experiments span distinct yet overlapping clusters. This distribution confirms the inclusion of both traditional and emerging alloy chemistries, enabling the training of models with broad generalization capacity.

The full data curation process is visualized in Figure 2c. Starting from 11,403 candidates, we applied a multi-stage filtering pipeline. Alloys were retained only if fabricated under as-cast conditions ensuring consistency in processing state, and tested at standardized conditions (room temperature, 283–311 K). Entries lacking reliable property measurements or complete composition data were excluded. After rigorous cleansing[31], we retained 837 high-fidelity as-cast alloy records for model training. Each alloy entry includes the fractional composition of up to 118 elements, a binary flag indicating casting condition, and experimentally reported values for four target mechanical properties, i.e., density, YS, UTS, and elongation. To focus the learning process on intrinsic composition–property relationships, we excluded processing and structural descriptors from model inputs.

To identify key composition–property trends, Figure 2d presents a dual association map. The left panel shows statistically significant links between elements and properties using Mantel tests, where edge thickness encodes association strength and colour indicates P-value. The right heatmap shows pairwise correlations among target properties and the overall composition vector. Notably, Fe, Ni, Mn, Mg, and Cr exhibit positive associations with YS and UTS, aligning with their established roles in strengthening mechanisms[5,32–34]. The property space of the curated dataset is further illustrated in Figure 2e–g, which provides critical insights into achievable performance limits and highlights regions where alloy discovery efforts can yield the greatest impact. In the strength–ductility map (Figure 2e), we observe the canonical trade-off: alloys with high UTS often exhibit limited elongation, while those with good ductility tend to have lower strength. Ti-based and Al-based systems demonstrate relatively balanced performance, suggesting their potential as promising platforms for optimization. However, the scarcity of alloys in the upper-right quadrant, i.e., high UTS and high elongation, indicates that few known compositions simultaneously achieve both mechanical robustness and ductility, underscoring a key target for inverse design. Figure 2f plots UTS against density, mapping the specific strength landscape. The red dashed line denotes the best specific-strength frontier (constant UTS/density), providing a visual benchmark for alloy efficiency. Most existing alloys lie well below this frontier, especially in the low-density regime (<4.5 g/cm³), revealing significant room for improvement. Designing new lightweight alloys that approach or surpass this frontier is a major focus of this study. Similarly, Figure 2g shows YS versus density, reinforcing the same pattern: very few alloys achieve both high YS and low density. The sparsity of data near the upper-left region (high strength, low mass) signals the opportunity space for MATAI-guided alloy design. Together, these maps quantify the current boundaries of achievable performance and reveal strategic regions for exploration. Specifically, the highlighted opportunity space, comprising high strength, high ductility, and low density, defines the multi-objective design target for the MATAI framework in the Ti alloy case study presented in Section 2.4.

## 2.3. Foundational Alloy Property Predictors

At the core of the MATAI framework is a set of foundational predictors trained to estimate key mechanical properties of alloys, including density, YS, UTS, and elongation, directly from elemental composition. These models act as fast, high-fidelity surrogates for real-world experiments, enabling efficient screening, optimization, and iterative feedback.

Each predictor is implemented as a multi-layer perceptron (MLP), which takes the fractional concentration of 118 elements as input and outputs the z-score of the target property. Compared to traditional ML techniques such as support vector machine (SVM)[35,36] and tree-based models like XGBoost[37,38], deep neural networks (DNNs)[39] learn complex, non-linear relationships without requiring handcrafted features, enabling accurate property estimation over broad composition spaces. To exploit inter-property correlations and improve sample efficiency, we adopt a multi-task learning (MTL)[40] strategy for the joint prediction of YS and UTS, which are highly positively correlated (Figure 2d). A shared MLP backbone extracts latent compositional features, followed by separate prediction heads for each property. To ensure that the model's predictions remain physically valid, we explicitly constrain the UTS must exceed or equal YS. Instead of predicting them independently, the model expresses UTS as the sum of YS and a non-negative offset, where the offset is generated through a softplus activation. This built-in physical prior prevents unphysical predictions (UTS < YS) and enhances both model robustness and physical fidelity.

All models are trained using the mean-squared error (MSE) loss, with property values standardized to stabilize training. We benchmark MLPs against classical models, including SVM and XGBoost[37], using 10-fold cross validation and grid-search-optimized hyperparameters. Figure 3a summarizes overall performance: MLPs consistently outperform baselines across all properties, achieving the highest $R^2$ values and lowest MAEs. The performance gap is especially pronounced for YS and UTS, where shallow models struggle with non-linearity and data sparsity. Boxplots in Figure 3b show distributional robustness across train–test splits. MLPs yield tighter and higher $R^2$ distributions, whereas SVM exhibits large variance and frequent degradation, particularly on strength metrics. XGBoost narrows this gap by combining multiple learners via gradient boosting[41], but still falls short of MLP. To visualize predictive fidelity, Figure 3c presents parity plots for the best-performing MLP models. Predictions align closely with ground-truth values, especially for density and tensile strength. While elongation shows slightly more scatter due to its inherent complexity, the overall trend remains strong, reinforcing the model's reliability.

To interpret the internal structure learned by the model, we project the latent composition embeddings into two dimensions for each property, shown in Figure 3d. Each point represents an alloy composition, with colour indicating the corresponding property value. These projections reveal physically meaningful groupings, connecting data-driven representations with metallurgical understanding. For density, Al-, Ti-, and Mg-based alloys cluster in the low-density regions, aligning with their well-known lightweight nature. In contrast, Fe- and HEA systems cluster at the high-density end due to the inclusion of heavier elements such as Cr, Co, Mo. For elongation, high values are predominantly localized within HEAs and Fe-based systems, reflecting their superior ductility and formability. On the other hand, alloys based on Al, Mg, and Ti tend to concentrate in the low-ductility regions, consistent with their more brittle behaviour in as-cast states. In terms of strength (YS and UTS), the embeddings highlight Ti-based and Fe-based alloys, as well as HEAs enriched with refractory elements such as Nb, Mo, Ta, and W, occupy the high-performance regions. These embeddings demonstrate that the model has internalized valid composition–property relationships without being explicitly told, supporting both predictive accuracy and interpretability.

Together, these results show that MATAI's foundational predictors not only deliver high-fidelity property estimation but also yield interpretable compositional landscapes. These maps guide downstream design by highlighting promising regions in the alloy space, ultimately enhancing the transparency and scientific utility of the framework.

**2.4. Generalist Alloy Designer: Ti-based alloy as a Case Study**

To identify high-performance alloys under realistic physical and manufacturing constraints, MATAI features a generalist alloy designer that formulates the inverse design task as a constrained, multi-objective optimization problem over compositional space. This component works in synergy with the foundational predictors to propose novel compositions that satisfy user-defined objectives such as maximizing strength and ductility while minimizing density, subject to physical and manufacturing feasibility.

We cast the alloy design problem as a COP defined by a tuple $\langle X, D, C, \tilde{C}, f \rangle$ where decision variables $X$ denotes the composition assigned to each eligible element, $D$ the discrete domains of allowable atomic percentages for each element, $C$ the set of hard constraints, $\tilde{C}$ the soft constraints, and $f$ the utility function to be optimized. The goal is to find an optimal assignment to each decision variable while satisfying all hard constraints. For the Ti-based alloy system, $X$ includes 15 major (e.g., Ti, Al, Sc, V, etc.) elements and 17 minor (e.g., B, C, Mn, etc.) elements, with each decision variable taking a value assignment from a domain discretized by a step size of 0.25%, resulting in a vast search space with size exceeding $10^{471}$.

Hard constraints ensure physical and practical feasibility, including: (1) the simplex constraint (total composition sums to 100%), (2) limiting the number of active elements to between 4 and 6, (3) capping the total content of minor element at 10%, and (4) enforcing that no non-Ti element exceeds the Ti fraction. In parallel, soft constraints encode user preferences based on different optimization directions. For instance, optimizing specific strength enforces upper bounds on density and lower bounds on elongation, while density-driven searches impose soft thresholds on strength and ductility.

The utility function combines the predicted target property with penalties for soft constraint violations, allowing flexible navigation of competing objectives. To solve the COP efficiently, MATAI adopts a bi-level optimization framework (Figure 4a). The upper level performs large neighbourhood search (LNS)[42,43] guided by the utility function, while the lower level employs constraint programming (CP)[25,26] via parallelized backtracking search[44] to ensure feasibility under hard constraints. This hybrid approach balances global exploration with strict adherence to physical and design rules. Specifically, in each iteration of optimization, MATAI first randomly destroys a proportion of the incumbent composition assignment to create a neighbourhood around the assignment. Then a parallelized backtracking search is employed to enumerate all possible feasible candidates within the neighbourhoods by repairing the discarded proportion. After that, candidates are evaluated by the foundational predictors and scored by the utility function, where the best candidate replaces the incumbent assignment according to the simulated annealing criterion[45,46].

We applied this generalist designer to the as-cast Ti-based alloy system, targeting the discovery of compositions that outperform commercial benchmarks such as TC4 alloy[47]. Besides hard constraints for physical and manufacturing feasibility, we also imposed soft constraints on ductility (elongation $\geqslant$ 5%) and density ($\leqslant$ 4.45 g/cm$^3$), reflecting practical requirements for lightweight structural alloys. Three distinct optimization objectives are explored in separate runs: (i) maximizing specific strength (UTS/density), (ii) maximizing the product of strength and ductility (UTS × elongation), and (iii) minimizing density while preserving their mechanical performances. In each iteration, the designer generates new candidate alloys, ranks them, and recommends top-performing compositions for experimental synthesis and testing. The experimental results are then fed back into the training set to refine both the predictors and constraints in a closed-loop fashion. This iterative, bi-level optimization procedure progressively narrows the search space and improves design quality across cycles. The anytime optimization trajectories[48] are visualized in Figure 4b. The background contours represent the model-estimated property landscape, while grey points are previously known alloys. The coloured paths trace how MATAI progressively improves the design across iterations. Notably, all paths move beyond the dense cluster of existing alloys to reach underexplored regions with higher predicted utility, demonstrating that MATAI does not merely interpolate existing knowledge, but actively explores and exploits promising, under-sampled regions of composition space to identify novel, high-performance alloys.

## 2.5 Iterative AI–Experiment Feedback Loop

To translate algorithmic predictions into real-world alloys, we integrated MATAI's generalist designer into a closed-loop experimental workflow. This iterative AI–experiment feedback loop bridges data-driven design with physical validation, enabling the system to not only test its hypotheses but also refine them over time. By cycling between candidate generation, experimental synthesis, and data reintegration, MATAI continuously improves its prediction accuracy and design capability.

In the first three iterations, we fabricated the top-ranked candidates directly proposed by the generalist designer. This early-phase strategy emphasized autonomous exploration and model bootstrapping. Ten candidates failed to be synthesized, likely due to processability issues. To incorporate this negative feedback, we assigned these samples zero-valued properties, enabling the model to implicitly learn to avoid similar infeasible compositions in the future. From the fourth iteration onward, a more selective strategy was adopted. We ranked the top 50 candidate compositions based on three composite metrics: low density, high specific strength, and high product between YS and elongation. Among these, candidates exhibiting the lowest densities were prioritized for synthesis. This strategy aimed to steer the system toward high-performance, lightweight alloys.

Figure 5 summarizes the evolution of experimental performance across the seven optimization cycle. A clear trend emerges: density values decrease steadily from the fourth to the seventh iteration, as shown in Fig. 5a, indicating successful exploration of lighter compositions. Simultaneously, YS exhibits substantial gains from 712 MPa to 1096 MPa, with several alloys surpassing 1,200 MPa in UTS (Fig. 5b). Ductility is largely maintained above 5% throughout the process, with a peak value of 9.7% observed (Fig. 5c). Among the best-performing candidates, the Ti-based alloy exhibit a compelling balance of properties, achieving a low density of 4.41 g/cm$^3$, high YS of 934 MPa, UTS of 1045 MPa, and ductility of 5.6% at room temperature, superior than the benchmark TC4 alloy[47]. As shown in Figure 2f–g, this alloy occupies a position near the frontier in both the UTS–density and YS–density plots. This placement highlights MATAI's ability to navigate previously unexplored, high-performance regions of composition space that balance multiple conflict objectives, thereby extending the current performance envelope of records.

To understand the microstructural origins of this performance, we conducted extensive characterization. Electron Backscatter Diffraction (EBSD) and Backscattered Electron (BSE) imaging (Fig. 5d) revealed a lath-like α microstructure interspersed with nanoscale β precipitates. These features are consistent with strengthening mechanisms known to operate in Ti-based alloys[49]. High-Angle Annular Dark-Field Scanning Transmission Electron Microscopy (HAADF-STEM) together with Energy Dispersive Spectroscopy (EDS) mappings (Fig. 5e) indicated these precipitates were enriched in Cr, Mn and V elements, while the surrounding α matrix consists predominantly of Ti and Al. These solute distributions suggest that the minor alloying elements act as β-phase stabilizers, promoting the formation of finely dispersed precipitates that can impede dislocation motion and strengthen the alloy. High-resolution TEM (HRTEM) and the associated fast Fourier transform (FFT) patterns confirmed a fully coherent interface between the α matrix and the β precipitates (Fig. 5f), indicative of good lattice matching and phase compatibility. This interfacial coherency reduces the interfacial defect density and contributes to the concurrent realization of high strength and ductility.

The results validate the effectiveness of MATAI's closed-loop learning and its ability to reconcile the strength and ductility in lightweight alloy design. By systematically integrating real-world performance data into model retraining, MATAI was able to refine its predictions and converge on compositions that not only outperform existing Ti-based benchmark but also redefine the performance boundaries within the design space[50].

## 3. Discussion

Designing high-performance alloys has traditionally relied on extensive expertise and costly, time-intensive experimental iterations to identify promising candidates, often requiring months to years trial-and-error. In this work, we present MATAI, a generalist ML framework for alloy property prediction and inverse design, which significantly streamlines and accelerates the discovery process by several orders of magnitude. By compiling diverse alloy data points across sources, MATAI first curates a high-quality, unified holistic database containing over 10,000 experimentally verified compositions. Based on the holistic database, MATAI trains a set of foundational predictors capable of estimating multiple alloy properties across broad composition spaces with high fidelity, serving as a fast surrogate for laborious and time-consuming real-world experiments. To navigate complex physical constraints and competing design objectives, MATAI's generalist designer formulates the alloy design task as a constrained optimization problem. It then employs a novel bi-level optimization framework to efficiently discover high-quality alloy compositions within minutes to hours. Finally, MATAI incorporates an iterative AI-experiment feedback loop, ensuring the continuous refinement by integrating results from real-world experimental validation on the proposed alloy candidates, discovering an alloy with a low density of 4.41 g/cm$^3$, while maintaining a high YS of 934 MPa with 5.6% ductility and ultimate tensile strength of 1045 MPa at room temperature.

Despite its performance and versatility, MATAI does not yet incorporate process or microstructural features, and its ability to adapt across diverse processing methods remains unexplored. In future, we will extend MATAI to integrate process parameters and structural information, coupled with deployment in autonomous laboratories to accelerate the closed-loop cycle of design, synthesis, and validation toward fully self-driving materials discovery.

**Declaration of competing interest**

The authors declare that they have no known competing financial interests or personal relationships that could have appeared to influence the work reported in this paper.

**Acknowledgements**

This work was supported by the National Research Foundation Singapore and DSO National Laboratories under the AI Singapore Programme (AISG Award No: AISG2-GC-2023-009). This work was supported in part by the AI2050 initiative at Schmidt Sciences (Grant G-25-68035).

**Author contributions**

Y.D., C.Z. and Y.L. contributed equally to this work. B.A., Z.L., X.W. and B.T. conceived and supervised the project. Y.D. designed the MATAI framework and developed search algorithms. C.Z. synthesized the samples and performed mechanical and structural characterizations. Y.L., P.Y. contributed to foundational predictors. Z.Z., Y.Lu, J.H. and Y.X. assisted with data analysis. Y.D, C.Z., Y.L. and B.T. co-wrote the manuscript. All authors discussed the results and contributed to the final manuscript.

**References**

1. Pollock, T. M. Alloy design for aircraft engines. *Nat. Mater.* **15**, 809–815 (2016).
2. Han, L. *et al.* Multifunctional high-entropy materials. *Nat. Rev. Mater.* **9**, 846–865 (2024).
3. Wei, S., Ma, Y. & Raabe, D. One step from oxides to sustainable bulk alloys. *Nature* **633**, 816–822 (2024).
4. Raabe, D. The Materials Science behind Sustainable Metals and Alloys. *Chem. Rev.* **123**, 2436–2608 (2023).
5. George, E. P., Raabe, D. & Ritchie, R. O. High-entropy alloys. *Nat. Rev. Mater.* **4**, 515–534 (2019).
6. Raabe, D., Tasan, C. C. & Olivetti, E. A. Strategies for improving the sustainability of structural metals. *Nature* **575**, 64–74 (2019).
7. Rao, Z., Springer, H., Ponge, D. & Li, Z. Combinatorial development of multicomponent Invar alloys via rapid alloy prototyping. *Materialia* **21**, 101326 (2022).

8. Pei, Z., Yin, J., Liaw, P. K. & Raabe, D. Toward the design of ultrahigh-entropy alloys via mining six million texts. *Nat. Commun.* **14**, 54 (2023).

9. Hu, Q.-M. & Yang, R. The endless search for better alloys. *Science* **378**, 26–27 (2022).

10. DebRoy, T., Mukherjee, T., Wei, H. L., Elmer, J. W. & Milewski, J. O. Metallurgy, mechanistic models and machine learning in metal printing. *Nat. Rev. Mater.* **6**, 48–68 (2020).

11. Sohail, Y. *et al.* Machine-learning design of ductile FeNiCoAlTa alloys with high strength. *Nature* **643**, 119–124 (2025).

12. Rao, Z. *et al.* Machine learning–enabled high-entropy alloy discovery. *Science* **378**, 78–85 (2022).

13. Zhou, Z. *et al.* Machine learning guided appraisal and exploration of phase design for high entropy alloys. *Npj Comput. Mater.* **5**, 128 (2019).

14. Yang, C. *et al.* A machine learning-based alloy design system to facilitate the rational design of high entropy alloys with enhanced hardness. *Acta Mater.* **222**, 117431 (2022).

15. Wang, Q. & Yao, Y. Harnessing machine learning for high-entropy alloy catalysis: a focus on adsorption energy prediction. *Npj Comput. Mater.* **11**, 91 (2025).

16. Wen, C. *et al.* Machine learning assisted design of high entropy alloys with desired property. *Acta Mater.* **170**, 109–117 (2019).

17. Zhang, Y. *et al.* Phase prediction in high entropy alloys with a rational selection of materials descriptors and machine learning models. *Acta Mater.* **185**, 528–539 (2020).

18. Harada, T. & Alba, E. Parallel Genetic Algorithms: A Useful Survey. *ACM Comput. Surv.* **53**, 1–39 (2021).

19. Andrieu, C., De Freitas, N., Doucet, A. & Jordan, M. I. An Introduction to MCMC for Machine Learning. *Mach. Learn.* **50**, 5–43 (2003).

20. Wang, W., Yang, Y. & Wu, F. Towards Data-And Knowledge-Driven AI: A Survey on Neuro-Symbolic Computing. *IEEE Trans. Pattern Anal. Mach. Intell.* **47**, 878–899 (2025).

21. Brailsford, S. C., Potts, C. N. & Smith, B. M. Constraint satisfaction problems: Algorithms and applications. *Eur. J. Oper. Res.* **119**, 557–581 (1999).

22. Freuder, E. C. & Wallace, R. J. Partial constraint satisfaction. *Artif. Intell.* **58**, 21–70 (1992).

23. Pirlot, M. General local search methods. *Eur. J. Oper. Res.* **92**, 493–511 (1996).

24. Hoos, H. H. & Stützle, T. Stochastic Local Search. in *Handbook of Approximation Algorithms and Metaheuristics, Second Edition* (ed. Gonzalez, T. F.) 297–307 (Chapman and Hall/CRC, 2018). doi:10.1201/9781351236423-17.

25. Dechter, R. *Constraint Processing*. (Morgan Kaufmann Publishers, San Francisco, 2003).

26. Rossi, F., Van Beek, P. & Walsh, T. *Handbook of Constraint Programming*. (Elsevier, Amsterdam, 2006).

27. Borg, C. K. H. *et al.* Expanded dataset of mechanical properties and observed phases of multi-principal element alloys. *Sci. Data* **7**, 430 (2020).

28. Ghorbani, M., Boley, M., Nakashima, P. N. H. & Birbilis, N. A machine learning approach for accelerated design of magnesium alloys. Part A: Alloy data and property space. *J. Magnes. Alloys* **11**, 3620–3633 (2023).

29. Ghorbani, M., Boley, M., Nakashima, P. N. H. & Birbilis, N. A machine learning approach for accelerated design of magnesium alloys. Part B: Regression and property prediction. *J. Magnes. Alloys* **11**, 4197–4205 (2023).

30. Jiang, L. *et al.* A rapid and effective method for alloy materials design via sample data transfer machine learning. *Npj Comput. Mater.* **9**, 26 (2023).

31. Li, K. *et al.* Exploiting redundancy in large materials datasets for efficient machine learning with less data. *Nat. Commun.* **14**, 7283 (2023).

32. Mohrbacher, H. & Kern, A. Nickel Alloying in Carbon Steel: Fundamentals and Applications. *Alloys* **2**, 1–28 (2023).

33. Yan, X. *et al.* Influence of the Mg content on the microstructure and mechanical properties of Al-xMg-2.0Si-0.6Mn alloy. *J. Mater. Res. Technol.* **23**, 3880–3891 (2023).

34. Wu, Y. *et al.* An overview of microstructure regulation treatment of Cu-Fe alloys to improve strength, conductivity, and electromagnetic shielding. *J. Alloys Compd.* **1002**, 175425 (2024).

35. Cortes, C. & Vapnik, V. Support-vector networks. *Mach. Learn.* **20**, 273–297 (1995).

36. Chang, C.-C. & Lin, C.-J. LIBSVM: A library for support vector machines. *ACM Trans. Intell. Syst. Technol.* **2**, 1–27 (2011).

37. Chen, T. & Guestrin, C. XGBoost: A Scalable Tree Boosting System. in *Proceedings of the 22nd ACM SIGKDD International Conference on Knowledge Discovery and Data Mining* 785–794 (ACM, San Francisco California USA, 2016). doi:10.1145/2939672.2939785.

38. James, G., Witten, D., Hastie, T., Tibshirani, R. & Taylor, J. Tree-Based Methods. in *An Introduction to Statistical Learning* 331–366 (Springer International Publishing, Cham, 2023). doi:10.1007/978-3-031-38747-0_8.

39. LeCun, Y., Bengio, Y. & Hinton, G. Deep learning. *Nature* **521**, 436–444 (2015).

40. Zhang, Y. & Yang, Q. A Survey on Multi-Task Learning. *IEEE Trans. Knowl. Data Eng.* **34**, 5586–5609 (2022).

41. Hastie, T., Tibshirani, R. & Friedman, J. *The Elements of Statistical Learning*. (Springer New York, New York, NY, 2009). doi:10.1007/978-0-387-84858-7.

42. Pisinger, D. & Ropke, S. Large Neighborhood Search. in *Handbook of Metaheuristics* (eds Gendreau, M. & Potvin, J.-Y.) vol. 272 99–127 (Springer International Publishing, Cham, 2019).

43. Shaw, P. Using Constraint Programming and Local Search Methods to Solve Vehicle Routing Problems. in *Principles and Practice of Constraint Programming* (eds Maher, M. & Puget, J.-F.) vol. 1520 417–431 (Springer Berlin Heidelberg, Berlin, Heidelberg, 1998).

44. Van Beek, P. Backtracking Search Algorithms. in *Foundations of Artificial Intelligence* vol. 2 85–134 (Elsevier, 2006).

45. Van Laarhoven, P. J. M. & Aarts, E. H. L. Simulated annealing. in *Simulated Annealing: Theory and Applications* 7–15 (Springer Netherlands, Dordrecht, 1987). doi:10.1007/978-94-015-7744-1_2.

46. Delahaye, D., Chaimatanan, S. & Mongeau, M. Simulated Annealing: From Basics to Applications. in *Handbook of Metaheuristics* (eds Gendreau, M. & Potvin, J.-Y.) vol. 272 1–35 (Springer International Publishing, Cham, 2019).

47. Oh, S.-T., Woo, K.-D., Kim, J.-H. & Kwak, S.-M. Effect of Retained β Phase on Mechanical Properties of Cast Ti-6Al-4V Alloy. *Mater. Trans.* **58**, 1145–1149 (2017).

48. Zilberstein, S. Using Anytime Algorithms in Intelligent Systems. *AI Mag.* **17**, 73–83 (1996).

49. Qu, Z. *et al.* High fatigue resistance in a titanium alloy via near-void-free 3D printing. *Nature* **626**, 999–1004 (2024).

50. Williams, J. C. & Starke, E. A. Progress in structural materials for aerospace systems11The Golden Jubilee Issue—Selected topics in Materials Science and Engineering: Past, Present and Future, edited by S. Suresh. *Acta Mater.* **51**, 5775–5799 (2003).

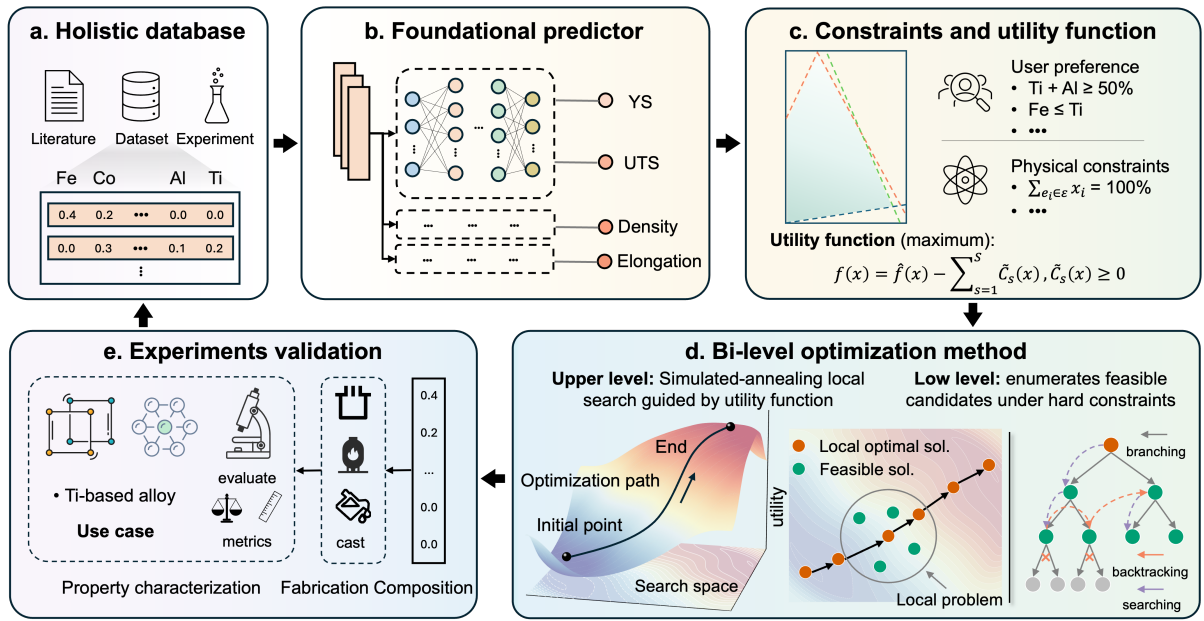

**Fig.1 | Overview of the MATAI framework for generalist alloy design. a,** Holistic database integrating curated literature, public datasets and in-house experiments, unified as composition–property records. **b,** Foundational predictors estimate YS, UTS, density and elongation from composition. **c,** Constraints and utility define feasible design space and user preferences. **d,** Bi-level optimization couples simulated-annealing local search (upper level) with a constraint-programming backtracking solver (lower level) to enumerate feasible candidates and rank them by utility. **e,** Experimental validation fabricates and tests recommended compositions, and returns results to refine the database and models.

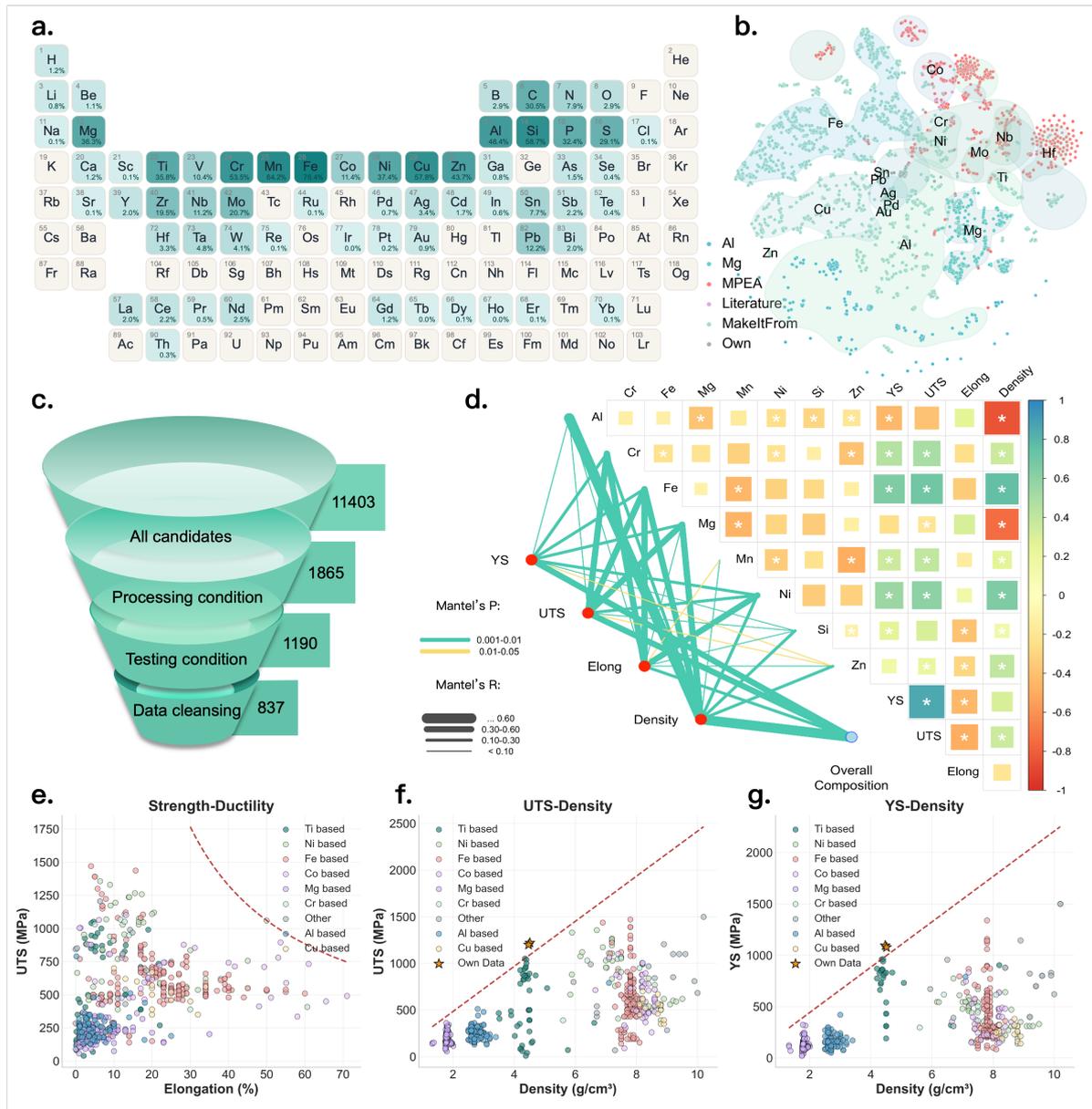

**Fig. 2 | Holistic alloy dataset, data curation pipeline, and composition–property analyses. a,** Periodic table showing elemental coverage of the compiled dataset; darker shading indicates higher frequency of element occurrence across compositions. **b,** Two-dimensional t-SNE embedding of all alloy compositions, coloured by data source: Al, Mg, MPEA, Literature, MakeItFrom, and Own. **c,** Data processing funnel showing progressive filtering from raw entries to high-fidelity as-cast records. **d,** Composition–property association analysis. Top-right: heat map showing Pearson correlation between properties and overall composition; asterisks indicate significance ($P < 0.05$). Left: Mantel test network linking elements to properties; edge colour and thickness reflect statistical strength and correlation magnitude, respectively. **e,** Strength–ductility plot (UTS vs. elongation), coloured by base element category. **f,** UTS versus density plot with dashed red line denoting the specific-strength frontier; AI-designed alloy from

this study lies near the frontier. **g,** YS versus density with the same reference; AI-designed alloy demonstrates simultaneous improvements in strength and lightweighting beyond existing data.

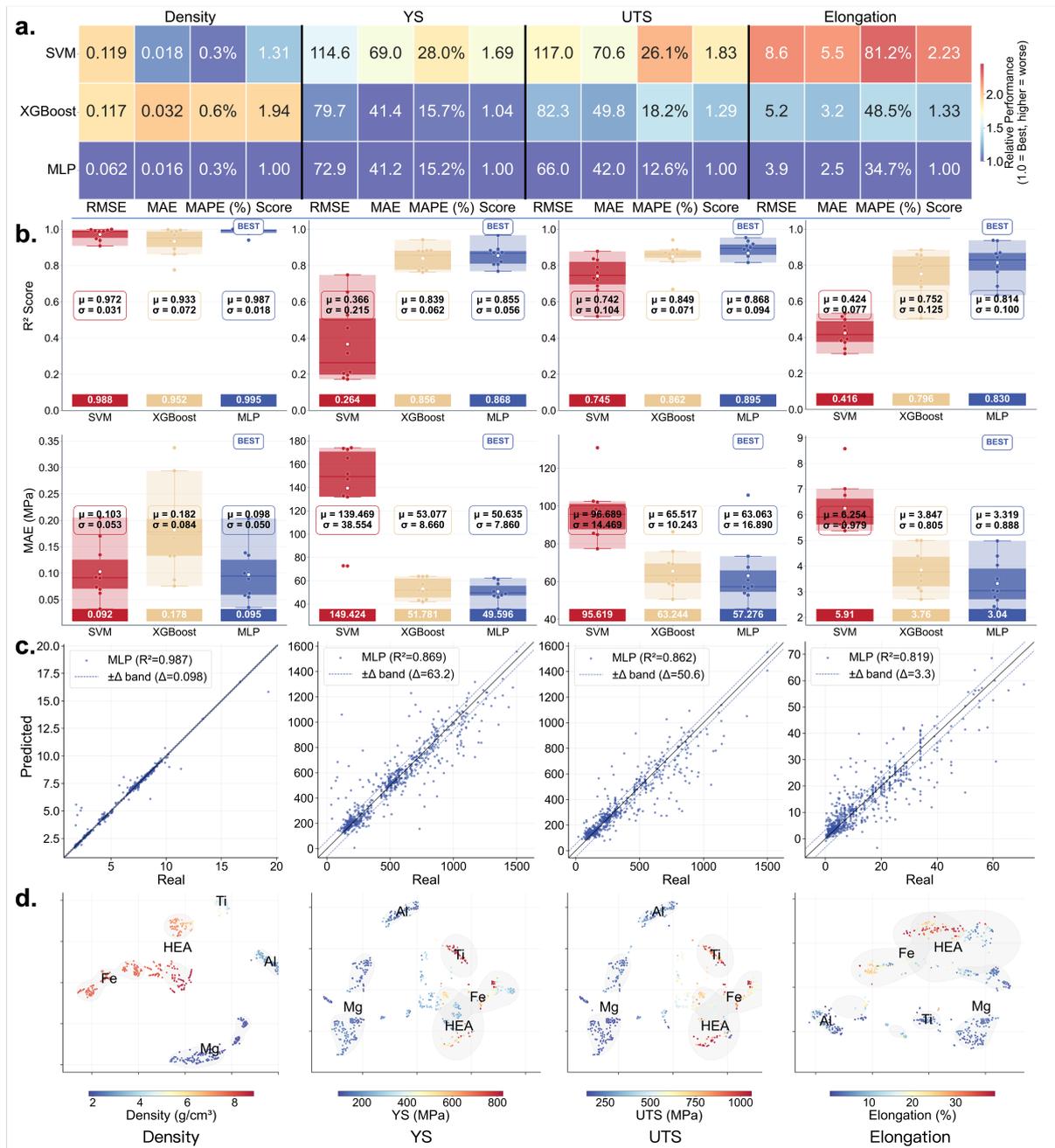

**Fig. 3 | Benchmarking and interpretation of foundational property predictors. a,** Performance summary of SVM, XGBoost, and MLP models across four targets: density ($\rho$), elongation ($\delta$), YS ($\sigma_y$), and UTS ($\sigma_{UTS}$) based on mean $R^2$ and MAE values from repeated train–test splits. Darker shading indicates better performance. **b,** Distribution of $R^2$ and MAE for each model and property. Boxes denote the interquartile range with the median line; whiskers show 1.5× IQR; points are individual splits of 10-fold cross validation. **c,** Parity plots for the best-performing MLP predictors for each target property. The solid line is y = x. **d,** Two-dimensional composition embeddings for density, elongation, YS, and UTS.

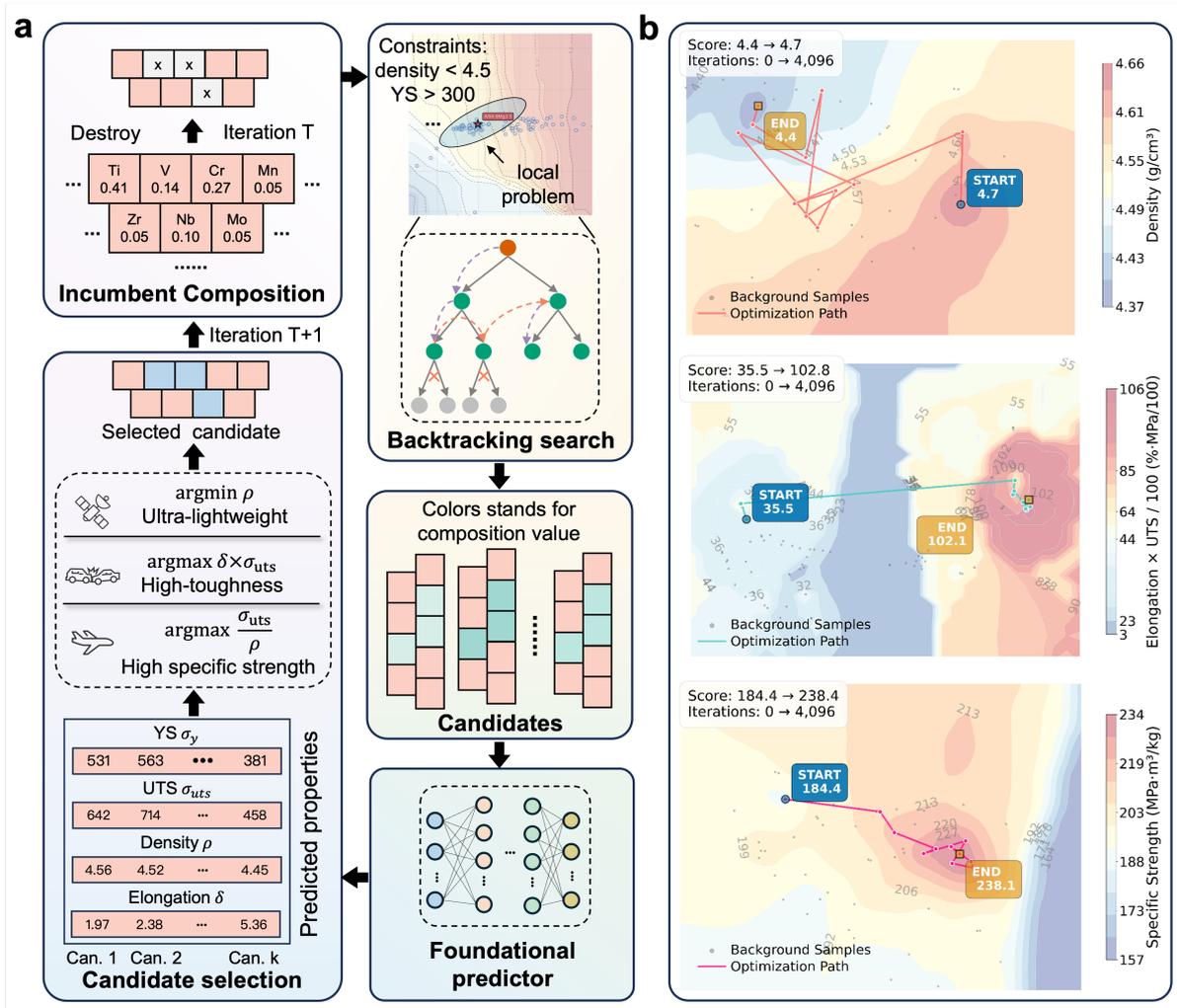

**Fig. 4 | Bi-level optimisation for task-driven alloy design. a,** Schematic of each optimisation iteration. From an incumbent composition, a destroy–repair move proposes candidate compositions. A backtracking search explores the feasible neighbourhood under hard constraints, guided by the foundational predictor. Candidates are then scored for task-specific objectives and the best is selected to become the new incumbent. **b,** Optimisation trajectories on learned composition-embedding landscapes for the three objectives. Background contours denote the property field estimated from the dataset; grey points are existing samples. Paths trace the sequence of proposed compositions from START to END, with the reported scores showing improvement over iterations.

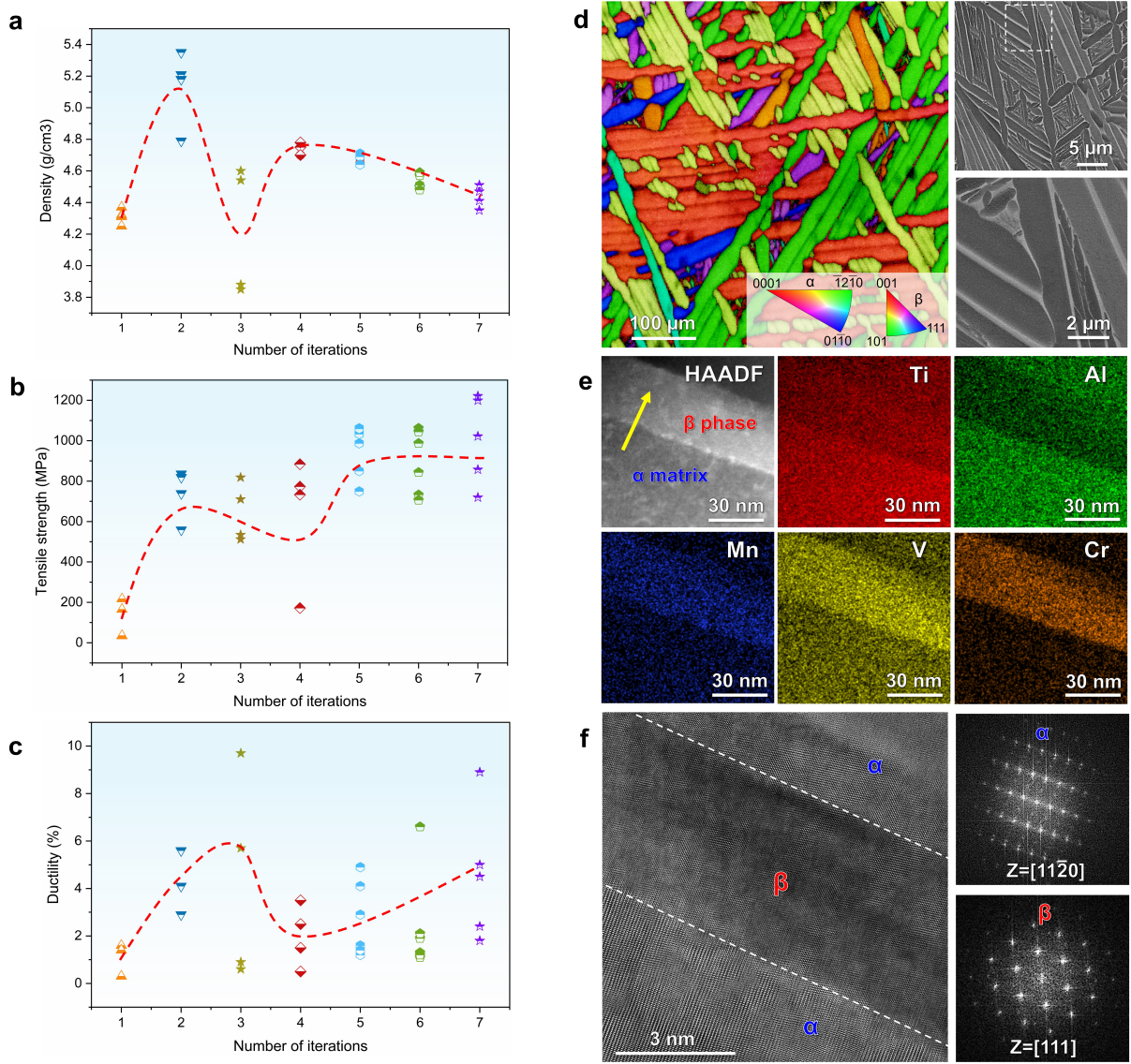

**Fig. 5 | Experimental validation and microstructural analysis of MATAI-designed Ti-based alloys. a-c,** Evolution of density, tensile strength, and ductility across seven AI–experiment iterations. Red dashed lines show general performance trends. **d,** The EBSD inverse pole figure (IPF) colour map of the as-cast Ti-based alloy; the right panels display BSE images at different magnifications. **e,** HAADF-STEM image and EDS mappings of the α-β region of the as-cast Ti-based alloy. **f,** High-resolution TEM image of the α-β region of the as-cast Ti-based alloy, with its corresponding FFT images displayed on the right.